# Training Massive Deep Neural Networks in a Smart Contract: A New Hope

Yin Yang
yin@yang.net

*Abstract*— Deep neural networks (DNNs) could be very useful in blockchain applications such as DeFi and NFT trading. However, training / running large-scale DNNs as part of a smart contract is infeasible on today's blockchain platforms, due to two fundamental design issues of these platforms. First, blockchains nowadays typically require that each node maintain the complete world state at any time, meaning that the node must execute all transactions in every block. This is prohibitively expensive for computationally intensive smart contracts involving DNNs. Second, existing blockchain platforms expect smart contract transactions to have deterministic, reproducible results and effects. In contrast, DNNs are usually trained / run lock-free on massively parallel computing devices such as GPUs, TPUs and / or computing clusters, which often do not yield deterministic results.

This paper proposes novel platform designs, collectively called *A New Hope* (*ANH*), that address the above issues. The main ideas are (i) computing-intensive smart contract transactions are only executed by nodes who need their results, or by specialized serviced providers, and (ii) a non-deterministic smart contract transaction leads to uncertain results, which can still be validated, though at a relatively high cost; specifically for DNNs, the validation cost can often be reduced by verifying properties of the results instead of their exact values. In addition, we discuss various implications of ANH, including its effects on token fungibility, sharding, private transactions, and the fundamental meaning of a smart contract.

## I. INTRODUCTION

Smart contracts, which are collections of executable computer code on a blockchain, have been criticized as "neither smart nor contracts" [9]. This paper aims to make smart contracts smarter by incorporating large-scale deep neural networks (DNNs) into the code, which has numerous potential applications. For instance, in decentralized finance (DeFi), a DNN might help detect abnormal token price movements, which could be part of a flash-loan attack [4]. A decentralized autonomous organization (DAO) [5] might trade tokens automatically with a DNN trained continually through reinforcement learning, *e.g.*, AlphaPorfolio [6]. A content creator might apply a generative adversarial network (GAN) [10] to generate visual art images, and subsequently tokenize them as non-fungible tokens (NFTs) tradable in a decentralized exchange. In particular, this paper aims to incorporate into smart contracts not only DNN inference, but training as well, since DNNs often need to be re-trained or fine-tuned with new data in many applications, *e.g.*, in reinforcement learning [18] and time series forecasting [19].

However, with today's blockchain platforms, training or running DNNs within a smart contract still appears to be a distant goal, for two main reasons. The first is cost. In Ethereum [8], for example, each smart contract instruction incurs a monetary cost called *gas*, which is about 5 GWei as of June 2021[1] (around $10^{-5}$ US dollars[2]). On the other hand, a large DNN is usually computationally intensive, leading to a prohibitively high gas cost. Specifically, Nvidia RTX 3090, a popular GPU for deep learning, can execute over 35 trillion floating-point operations (TFLOPs) per second[3]; even so, training a larger DNN on such a GPU can take hours or days. In terms of gas cost, such a process would easily burn millions of dollars.

Can we make smart contracts cheaper by artificially lowering gas prices? For example, Binance Smart Chain (BSC)[4] gained popularity rapidly in 2021, in part thanks to its significantly lower (often by a factor of 10x) gas cost compared to Ethereum. However, there is a fundamental limit on how cheap gas can be: that on current blockchain platforms, each node is required to execute all transactions in all blocks, and maintain the entire world state at all times, which includes account balances and smart contract storage of all accounts in the entire blockchain. Consequently, a node can only finish processing a block (say, block *b*), after the node completes executing all transactions in *b*. If the gas price is too low, and the block *b* contains a transaction with intensive computations (*e.g.*, training a large DNN) that takes much time (*e.g.*, days) to execute, then the processing time of block *b* is necessarily long, leading to prohibitively high latency for processing and validating block *b*. Partly due to this reason, platforms such as Ethereum explicitly limit the total amount of gas consumed (independent of gas price) within one block.

The second major hurdle for training / running a DNN within a smart contract transaction is that most blockchain platforms implicitly assume that the results of a transaction (including return values, token transfers, and modifications of storage state) are deterministic and reproducible. For example, in Ethereum, the hash value of block *b* covers the root hash of the global state tries [22], which is computed based on the account balances and smart contract storage states after executing all transactions in block *b*. When verifying a block, a validator (i) re-executes all transactions in *b*, (ii) re-computes the block hash based on the updated world state, and then (iii) checks whether this hash matches the one included in *b*. On the other hand, many algorithms involved in DNN training do not output deterministic results. For instance, the documentation of PyTorch, a popular Deep Learning library, lists nondeterministic functions[5]. As of June 2021, this list includes commonly used functions such as tensor interpolation (*e.g.*, for resizing an image), max / average pooling (commonly used in convolutional neural networks), and CTC loss [11] (for learning sequence alignment). Further, note that a "deterministic" function in PyTorch only guarantees reproducibility when the same code is run on the same hardware; in other words, different types of GPUs can still lead to different results. Lastly, some parallel DNN training algorithms such as Hogwild! [15] are specifically designed to sacrifice result determinism to gain high efficiency.

---

[1] *https://ethereumprice.org/gas/*

[2] *https://nomics.com/markets/gwei-gwei/usd-united-states-dollar*

[3] *https://www.nvidia.com/en-eu/geforce/graphics-cards/30-series/rtx-3090/*

[4] *https://www.binance.org/en#smartChain*

Is it possible to validate computer programs with long running time and non-deterministic results without centralized trusted authorities? This paper draws inspiration from common practices in the machine learning field, which has been advancing rapidly with a high degree of transparency. In particular, it is common for machine learning researchers to release source code, training data, and pre-trained models in a public repository, *e.g.*, Github (github.com), along with a published paper. Other researchers interested in the same subject then download the code and data, run the code on the dataset to train a new model, and verify that the test accuracy of the new model is consistent with that obtained by the pre-trained model and the claims of the paper. Unlike validators in the blockchain world, no machine learning researcher would attempt to run all versions of all code ever released in all repositories, for the pointless purpose of "maintaining the world state". Instead, *code is run on a need-to-run basis*.

Meanwhile, the fact that the results of many DNN training algorithms are not exactly reproducible is often just a minor inconvenience. In particular, when a researcher validates the release source code, dataset, and pre-trained model of a paper, the validation process focuses on obtaining comparable performance (*e.g.*, inference accuracy on a test dataset) as the released results of the code (*i.e.*, the pre-trained model), rather than reproducing an exact, bit-to-bit copy of the results. In other words, *the validation process runs another program to verify the properties of execution results to establish correctness*, which bypasses the irreproducible result problem.

As an extreme case, there exist massive DNNs with billions or even trillions of parameters, whose training is essentially limited to a few so-called "big-compute" organizations with vast amounts of computing power. For instance, the training of GPT-3 [3] was performed in a large data center, and cost millions of dollars. Further, the training data could be proprietary in some cases, *e.g.*, Google's internal image collection JFT-300M [17]. Even for such cases, the machine learning community still managed to validate the released pre-trained models (*e.g.*, a Transformer-based language model [7]) by verifying their effectiveness when fine-tuned on another task (*e.g.*, natural language inference [21]) with a dataset not involved in the original training process of the DNN. *In essence, here the validation of a program's outputs verifies the utility of these outputs in the target applications*. In this sense, the original program that obtained these results could be treated as a black box. For instance, in the case of GPT-3 [3], the authors did not release their pre-trained DNN model; instead, they allowed external researchers to interact with the model through a web interface, as a limited form of validation.

Based on the above observations, this paper proposes a set of fundamentally new blockchain platform designs, collectively called *A New Hope* (*ANH*), to enable the integration of large-scale DNNs into smart contracts. Specifically, ANH has the following characteristics:
- Validators of new blocks do not execute the transactions therein.
- Transaction execution is on-demand, possibly through a service provider called an on-chain accountant.
- Smart contract transactions are allowed to have non-deterministic results, which are verified through a special validation mechanism that may involve invoking other smart contracts.

In the following, Section II focuses on handling compute-intensive smart contracts that still yield deterministic results. Section III (which will appear in the next version of this pre-print) deals with smart contracts with non-deterministic results.

## II. HANDLING COMPUTING-INTENSIVE SMART CONTRACTS

This section focuses on enabling computing-intensive operations such as DNN training in smart contracts. In this section, all transactions are assumed to be deterministic; transactions with non-deterministic results are handled later in Section III.

### A. Validating Blocks without Executing Transactions

In existing blockchain systems, a block consists of a sequence of transactions and additional verification information such as signatures and hash values. For example, a block (say, $b_i$) in Ethereum [22] contains the following contents:
- Block number (*i.e.*, $i$) and the hash value in the previous block (i.e., $b_{i-1}$).
- An ordered list of transactions, denoted as $b_i.txs$.
- Values required by the consensus protocol, which is orthogonal to this paper.
- Root hash of the world state (denoted as $b_i.ws$) after executing all transactions in $b_i.txs$. Specifically, $b_i.ws$ covers the balance and nonce (*i.e.*, the number of transactions initiated) of every account, as well as the code and storage states of each smart contract, on the entire blockchain.

---
[5]*https://pytorch.org/docs/stable/generated/torch.use_deterministic_al gorithms.html#torch.use_deterministic_algorithms*.

Among the above items, only the last one, i.e., $b_i.ws$, poses a fundamental limit on the maximum computational cost of a smart contract transaction. Specifically, to generate $b_i.ws$, the node that creates (or verifies) block $b_i$ must (i) maintain the world state prior to $b_i$ and (ii) run all transactions in $b_i.txs$. Due to (ii), creating a new block $b_i$ takes at least as much time as running all transactions in $b_i.txs$, which can be long when a transaction in $b_i.txs$ contains expensive operations, e.g., training a DNN. Meanwhile, the creation of different blocks cannot be pipelined to hide the cost, since the creation of the next block $b_{i+1}$ can only be done once $b_i$ is confirmed, due to the above requirement (i).

To address the above problem, the proposed approach ANH takes a bold move of removing $b_i.ws$ altogether, for any $i>0$, meaning that with a single exception of the genesis block $b_0$, no block contains information about the world state. Further, a block can be formed as soon as its creator node gathers a list of transactions $b_i.txs$; a validator of $b_i$ simply verifies the signature of each transaction in $b_i.txs$, and completes the consensus process (orthogonal to this paper). In other words, neither the block creator nor the validators are required to run any transaction in $b_i.txs$.

**Double-spending attacks.** An immediate question about the above approach is: is the blockchain still robust against double-spending attacks? To answer this, let's examine a simple double-spending scenario: a user Alice with account $x$ and balance $\delta_x$ initiates two properly signed transfers: one to Bob and another to Carol, both with all the money in Alice's account, i.e., $\delta_x$. Without maintaining the world state, block creators and validators may not know Alice's account balance $\delta_x$ [6]; consequently, both transfers are added to the blockchain. Does this mean Alice has successfully performed a double-spending attack?

The answer, fortunately, is no. The reason is that although the blockchain contains both transactions Alice→Bob and Alice→Carol, due to the consensus protocol of the blockchain, there is a unique order of these two transactions. Hence, when someone (clarified later in Section II-B) eventually executes these two transactions, the later one of them inevitably fails due to insufficient balance[7], preventing the double-spending.

In general, *given an initial world state, a sequence of deterministic transactions with a fixed order uniquely determines the world state after executing these transactions*, which can be trivially proved by induction. In fact, this is commonly performed in log-based incremental data warehouse replication and Change Data Capture (CDC), where a log file records a sequence of transactions [16]. Since the blockchain's consensus protocol establishes a unique transaction order, there is no ambiguity of the final world state $b_i.ws$ after a block $b_i$, even though the block creator / validators have not explicitly computed $b_i.ws$.

**Observing transaction results.** One way to understand the above approach is that the world state $b_i.ws$ in fact always exists for every block $b_i$, just that no one has yet *observed* $b_i.ws$ by executing transactions in $b_i.txs$. Given sufficient time, any node can eventually observe $b_i.ws$ by executing all transactions that $b_i.ws$ depends upon. In other words, each transaction $tx$ committed to the blockchain (i.e., passing consensus) is considered *already happened* and has permanently and irreversibly altered the world state; its specific effects on the world states are only observed later on, when $tx$ is executed (detailed in Section II-B).

For instance, consider a DeFi lender smart contract with an automatic loan approval function that employs a DNN to decide on whether or not to approve the transaction sender's loan application. Now, imagine that thousands (depends on the block capacity) of loan applications are committed to the blockchain in a single block $b$. As soon as $b$ passes consensus (which is fast since none of the transactions are executed), the decision for each of the loan applications has already been made by the DNN, though at this point, no user knows these decisions. After that, each loan applicant can observe her result, by sequentially executing transactions in $b$ till reaching her own. Note that the observation process can take far longer than the creation of block $b$, due to the computationally intensive DNN.

**Challenges.** The above platform design in ANH, i.e., blocks of transactions are added to the blockchain before their execution, leads to two major challenges. The first is how to perform an on-chain to off-chain (O2O) transaction. For instance, consider that Alice transfers tokens to Bob in a transaction (denoted as $tx_{ab}$), in exchange for real-world (i.e., off-chain) goods or services. Assume that the $tx_{ab}$ is successfully added to the blockchain, say, in block $b_i$. Then, how does Bob know that he has indeed received the tokens, i.e., $tx_{ab}$ is not rolled back? Intuitively, Bob needs to observe his account balance after $tx_{ab}$, which can be computational intensive even though $tx_{ab}$ is a simple token transfer, because determining whether Alice's account has sufficient balance for $tx_{ab}$ can be non-trivial, especially when Alice's tokens come from complex smart contract (e.g., DeFi) transactions.

Another challenge is that an adversary can launch a *denial-of-service* (*DoS*) attack on the blockchain platform, by flooding the network with invalid transactions. An example of such attack is shown in the following algorithm.

---

*Algorithm 1: Tx-DoS*
1. Repeat (infinite loop):
2.     Generate a pair of keys ($k_{pub}$, $k_{priv}$) for a new account.
3.     Use the private key $k_{priv}$ to sign a new transaction $tx$ that transfers tokens to an arbitrary recipient.
4.     Submit the signed $tx$ to the blockchain platform.

---

In the above algorithm, the transaction $tx$ attempts to transfer tokens from a new account with zero balance, which is an invalid operation and has no effect on the world state. However, without explicitly maintaining the world states, the validators do not have account balance information, and, thus, must accept these transactions and put them in new blocks. Consequently, algorithm *Tx-DoS* leads to a flood of invalid transactions, which increases network transmissions and slows down the entire blockchain.

---

[6] Strictly speaking, the proposed approach ANH does require block creators and validators to verify that Alice has sufficient balance to pay the transaction fee. This is elaborated in Section II-D.

[7] When the execution of a transaction encounters an invalid operation, e.g., a transfer without sufficient balance, the whole transaction is rolled back, reversing all its modifications of the world state. This is a common practice in major blockchain platforms such as Ethereum.

In the following subsections, we address these challenges with novel platform designs in ANH.

*B. Transaction Indexing and Selective Execution*

As described in the previous subsection, in ANH, transactions are added to the blockchain pre-execution. When transactions contain expensive operations such as DNNs, maintaining the entire world state in real-time can become prohibitively expensive. Hence, ANH adopts a lazy transaction execution strategy: *that a transaction is executed only when its results are needed*. In particular, a user can investigate the value a given state $s$ at a given time (*e.g.*, $s \in b_i.ws$ for a given block $b_i$), by selectively executing transactions relevant to $s$. To facilitate fast retrievals of such transactions, the transactions can be indexed by an inverted file [1] that facilitates efficient state-to-transactions lookup, akin to how search engines index web pages on the Internet for fast keyword-based retrieval.

In our previous example that Alice commits a transaction $tx_{ab}$ to the blockchain transferring tokens to Bob, Bob would naturally want to check his own account balance after $tx_{ab}$. Ideally, all transactions on the blockchain are indexed by senders and receivers, so that Bob can quickly retrieve transactions relevant to his account, and executes only these to find his account balance. Suppose that Bob never submits computationally expensive smart contract transactions, then, ideally it should be relatively cheap for him to check his balance. Unfortunately, this is not true in the presence of adversaries: as we explain below, a simple query such as Bob checking his own account balance can be surprisingly expensive to answer.

**Execution DoS attacks.** Consider the following function, as part of a smart contract:

*Algorithm 2: Exec-DoS*
1. Perform some computationally expensive operations.
2. Randomly choose an account $x$ with a random seed obtained from the result of Line 1.
3. Transfer a small amount of tokens (*e.g.*, 1 Wei) from the parent smart contract account to account $x$.

A transaction calling the above smart contract function *Exec-DoS* can, in theory, modify the account balance of any user (and any smart contract) on the blockchain. In other words, the result of every account balance query depends on the execution of *Exec-DoS*, which involves computationally expensive operations (Line 1). Essentially, such transactions are a form of denial-of-service (DoS) attack of the entire blockchain platform, that significantly increases the cost of account balance queries.

To defend against *Exec-DoS*, a possible countermeasure is to impose financial penalties on accounts for initiating transactions invoking such a contract. For example, the blockchain platform could require that each transaction $tx$ explicitly declare the accounts and / or smart contract storage states that can be possibly affected by $tx$ [23], and charges a fee (elaborated in Section II-D) for each of them. Meanwhile, modifying an undeclared account is treated as an invalid operation, and any transaction with such an operation is rolled back. While such measures alleviate the problem of generic execution DoS attacks, it does not prevent a targeted attack on a specific account. Continuing our example, the adversary could initiate a transaction calling a contract similar to the following one, to increase the cost of account balance queries of a target victim, *e.g.*, Bob.

*Algorithm 3: Targeted-Exec-DoS($x$)*
// Input: $x$: an account in the blockchain.
1. Perform computationally expensive operations.
2. Randomly choose a small quantity $q$.
3. Transfer $q$ tokens to account $x$.

*Targeted-Exec-DoS* works against any given account $x$, making an account balance query computationally expensive for $x$ without the cooperation of $x$. Defending against *Targeted-Exec-DoS* is not easy. Instead, the proposed ANH platform takes a different approach: that the user is recommended to only perform *minimal accounting queries* that retrieve only the needed information (which does not depend upon *Targeted-Exec-DoS*) from the blockchain, explained below.

**Minimal accounting queries.** In our running example, observe that Bob, in fact, needs neither his exact account balance after $tx_{ab}$, nor Alice's exact account balance right before $tx_{ab}$. Instead, all Bob needs to know, is a single bit of information: whether or not Alice's account balance exceeds the amount transferred in transaction $tx_{ab}$, right before $tx_{ab}$, which is referred to as a *minimal accounting query* (*MAQ*) Intuitively, if Alice's account balance turns out to be insufficient for $tx_{ab}$, then, $tx_{ab}$ is rolled back, failing the token transfer; in this case, Bob should refuse to deliver the purchased goods or services. Conversely, if Alice's account has sufficient balance, then the transfer in $tx_{ab}$ irreversibly succeeds, meaning that Bob is obliged to deliver the purchased goods or services.

In general, the MAQ depends on (i) the blockchain application setting, and (ii) how the blockchain interfaces with the physical world. For instance, it is possible that Alice's transaction leads the modification of a smart contract storage state (*e.g.*, "vehicle insurance claim approved") that triggers Bob to perform an activity (*e.g.*, repairing a damaged car). The choice of the MAQ in such cases is an important part of the blockchain application design to make the application viable, which is decided by the app developer. In the following, we focus on the case of simple token transfer such as $tx_{ab}$ in our running example.

**Income and expenses.** It remains to clarify how Bob can efficiently perform an MAQ to check whether Alice has enough funds for the transfer transaction $tx_{ab}$. For this purpose, we first define two concepts: *expense* and *income* of a given account. Note that in this subsection, we assume that both Alice and Bob are user accounts. The cases for smart contract accounts are discussed later in Section II-E.

*Definition 1 (Expense).* Any transaction $tx$ sent from account $x$ that carries $v$ tokens represents an expense of amount $v$ w.r.t. account $x$.

*Definition 2 (Income).* Given a transaction $tx$ and an account $x$, let $\delta_x$ and $\delta'_x$ be the account balance of $x$ before and after $tx$, respectively. Then, there are two cases in which $tx$ represents an income of account $x$.

- Case 1: $tx$ is sent from another account $y \neq x$, and $\delta_x \neq \delta'_x$. In this case, $tx$ represents an income of amount $\delta'_x - \delta_x$ w.r.t. account $x$.
- Case 2: $tx$ is sent from $x$ carrying $v$ tokens, and $\delta_x - v \neq \delta'_x$. In this case, $tx$ represents an income of amount $\delta'_x - (\delta_x - v)$ w.r.t. account $x$.

With the above definitions, the balance of a user account $x$ is clearly equal to its total income subtracted by its total expenses. Note that unlike the concept of debit and credit in traditional accounting, in our context, a transaction $tx$ can represent *both* an expense and an income w.r.t. an account $x$. For instance, $x$ might send out tokens (which represents an expense of $x$) to a smart contract account $y$ as part of a function call, which subsequently triggers an operation (which can be from $y$, or another smart contract) that sends tokens back to $x$ (which represents an income of $x$). A special case concerns the payment of transaction fees, which is explained further in Section II-D.

The above definitions of expense and income ensure an important property: that *it is easy to compute the total expenses of a user account*. Specifically, all transactions on the blockchain can be stored in a database, indexed by the sender account, with a *value* column representing the amount of tokens carried in each transaction. Then, the total expense of a given account $x$ can be computed in a single SELECT query:

SELECT SUM(value) FROM blockchain WHERE sender=$x$

Clearly, the above query can be augmented to express a more fine-grained query, *e.g.*, the total expense of account $x$ right before a given transaction $tx$. Details are omitted for brevity. The total expense of a smart contract account, on the other hand, can be complex, and we defer the discussion to Section II-E.

The exact total income of an account, however, can be expensive to compute, for two reasons. First, the income may come from a computationally intensive smart contract function call (*e.g.*, a DNN, fine-tuned on new data, decides to send tokens to account $x$), or a sequence of such function calls. In fact, the two above-mentioned attacks *Exec-DoS* (Algorithm 2) and *Targeted-Exec-DoS* (Algorithm 3) are examples of such computationally expensive incomes. Second, the income transaction itself $tx$ might involve invalid operations and get rolled back, *e.g.*, $tx$ may be sent from an account $y$ with insufficient balance. In general, an income transaction may have far reaching dependencies and require the execution of a large number of transactions, possibly with a tremendous amount of computation. Hence, obtaining the exact total income of an account can be rather difficult.

**Account balance lower bounding.** To address the above problem, ANH recommends computing a lower bound of an account's total income with selected income transactions. Intuitively, the DoS attacks in Algorithm 2 and 3 can only affect an account $x$ by increasing its balance; thus, when computing a lower bound of $x$'s balance instead of its exact value, one can simply ignore such attacks.

Formally, we have the following lemma.

*Lemma 1 (account balance lower bound).* Given an account $x$ and a set of its income transactions $\Theta$, let $P_\Theta$ be the total income of $x$ from all transactions in $\Theta$. Let $Q$ be the total expenses of $x$, and $X_0$ be the initial token balance of $x$ specified in the genesis block $b_0$. Then, $X_0 + P_\Theta - Q$ is a lower bound of the account balance of $x$.

The proof of Lemma 1 is omitted for brevity, which follows directly from the definitions of income (Definition 1) and expenses (Definition 2).

Continuing our running example, Alice selects a set $\Theta$ of her income transactions, and presents $\Theta$ to Bob. The latter then executes transactions in $\Theta$, obtaining Alice's total income $P_\Theta$ represented by transactions in $\Theta$. If this total income, combined with Alice's initial balance at genesis (if any), and subtracted by Alice's total expenses (which can be retrieved efficiently as explained before), exceeds the value of $tx_{ab}$, then Bob can rest assured that the token transfer in $tx_{ab}$ has been successfully fulfilled. This process is summarized in the algorithm *Pay* below, where $x$ is Alice and $y$ is Bob in our running example.

---

*Algorithm 4: Pay* ($x$, $y$, $amt$)

// Inputs: $x$, $y$: sender and recipient accounts.
//          $amt$: amount to tokens that $x$ sends to $y$.
1. $x$ submits a transaction $tx$ transferring $amt$ tokens to $y$.
2. $x$ present to $y$ a list of income transactions $\Theta$.
3. $y$ queries the blockchain to obtain $x$'s total expenses $Q$ prior to $tx$, and its initial balance $X_0$ in the genesis block.
4. $y$ computes the results of transactions in $\Theta$ to determine whether $X_0 + P_\Theta - Q \geq amt$.
5. If $P_\Theta - Q \geq amt$, then:
6.    $y$ accepts $tx$, and performs the required real-world actions in exchange to the token transfer $tx$.
7. Else: $y$ rejects $tx$, and refuses to perform the corresponding real-world actions in exchange to $tx$.

---

Note that in Line 4 of the above algorithm, the recipient $y$ aims to determine whether $X_0 + P_\Theta - Q \geq amt$ (Lemma 1), which does not necessarily involve obtaining the exact value of $P_\Theta$. The fact that $P_\Theta$ can be lower-bounded is an opportunity for further optimization.

The *Pay* process is clearly robust to the attacks *Exec-DoS* (Algorithm 2) and *Targeted-Exec-DoS* (Algorithm 3), as long as these transactions are not added to the set $\Theta$, which is chosen by the sender $x$ (*e.g.*, Alice). Further, note that the use of a selected set $\Theta$ ensures that transactions are executed on a need-to-execute basis, which is a major goal of ANH.

It remains to clarify (i) how the sender (Alice) selects the income transaction set $\Theta$, and (ii) how the recipient (Bob) verifies that $P_\Theta - Q \geq amt$ with the minimum computations. We call these operations *computational accounting*, which can be done through professional services called *on-chain accountants*. In the next subsection, we detail the computational accounting process, and how on-chain accounts help in this process.

## C. Computational Accounting and On-Chain Accountants

**Choosing $\Theta$ by income cost.** First, we clarify the selection of the income transaction set $\Theta$, which is based on the concept of *income cost*, defined below.

*Definition 3 (Income Cost).* Given a transaction $tx$ that represents an income w.r.t. account $x$, the income cost $C_{tx}$ for $tx$ is the amount of computations (measured by gas consumption) required to determine the specific amount of income $v$ w.r.t. account $x$, as a result of $tx$. Transaction $tx$ is a *zero-cost income* w.r.t. $x$, iff. $C_{tx}=0$.

Note that in the above definition, the amount of income $v$ can be zero, *e.g.*, when $tx$ is rolled back. The income $v$, by definition, cannot be negative, since all outgoing token transfers are considered as expenses, including transaction fee payment, elaborated in Section II-D. The concept that each income has a cost resembles the real-world situation that incomes are taxed, which is discussed in Section II-E.

Once a user (*e.g.*, Alice) knows the income cost of each of her income transactions, the selection of set $\Theta$ reduces to a standard knapsack problem, *i.e.*, choosing the set of incomes with the minimum total cost that satisfies the inequality $X_0+P_\Theta-Q\geq amt$ in Lemma 1. Although the knapsack problem is NP-hard, it is known to have a fully polynomial-time approximation scheme [20], meaning that its optimal solution can be effectively and efficiently approximated.

**Calculating income costs.** Next, we focus on the calculation of income costs, which can itself become rather complicated and expensive. Figure 1 illustrates three examples of income cost. In the first example (Figure 1a), Alice had received 100 tokens from another account Carol in a transfer transaction $tx_{ca}$; Carol has an initial balance of 1 million, specified in the genesis block $b_0$, which is more than sufficient to cover all her expenses including $tx_{ca}$. Then, $tx_{ca}$ is a zero-cost income w.r.t. Alice, since the verifier (say, Bob) does not need to execute any smart contract code to confirm the value of $tx_{ca}$ (*i.e.*, 100, which is explicitly stated in $tx_{ca}$) or that the sender (Carol) has sufficient funds to pay Alice through $tx_{ca}$.

In the second example (Figure 1b), instead of a direct token transfer, Alice's rich friend Carol creates a smart contract account $y$, transfers 1,000 tokens to $y$, and then submits a smart contract transaction $tx_{ya}$ that triggers the transfer of tokens from $y$ to Alice. Then, $tx_{ya}$ represents an income to Alice of value 100, and its income cost equals the gas consumption of $tx_{ya}$.

In the last example (Figure 1c), there is another user Dave, who made a simple transfer of 100 tokens to Alice via transaction $tx_{da}$. Dave's income comes from a previous smart contract transaction $tx_{yd}$ which sent tokens from a smart contract $y$ set up and funded by Carol, similar to the second example. Then, the income cost of $tx_{da}$ is the gas consumption of $tx_{yd}$, because (i) $tx_{da}$ is a simple token transfer that consumes no gas, and (ii) to confirm the validity of $tx_{da}$, the verifier Bob needs to confirm the income of Dave, which involves the execution of $tx_{yd}$.

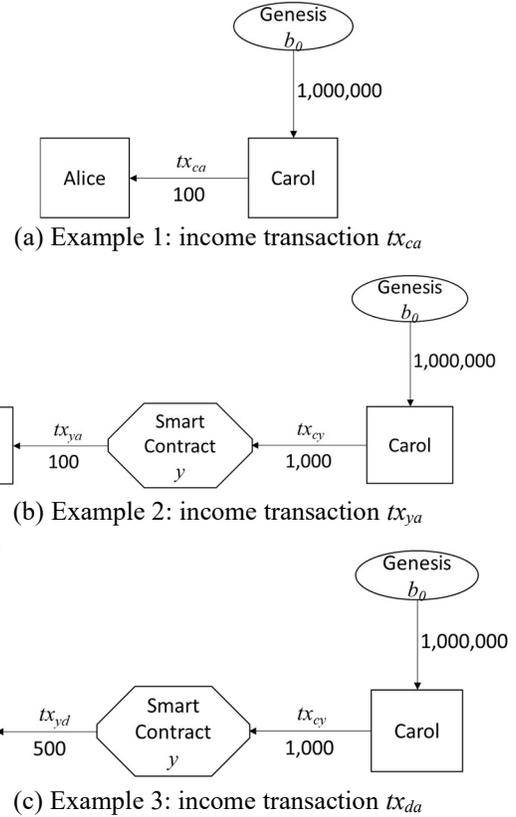

(a) Example 1: income transaction $tx_{ca}$

(b) Example 2: income transaction $tx_{ya}$

(c) Example 3: income transaction $tx_{da}$

**Figure 1:** Examples of income cost.

In general, the income cost of a transaction $tx$ is determined by the cost of computing the *provenance* of the world states that $tx$ depends upon, including account balances (which might be lower-bounded to reduce computations) and smart contract storage states (which usually need to be computed exactly). When $tx$ has complex dependencies, calculating its income cost can be computationally intensive.

Fortunately, a user often already knows the cost for at least some of her income transactions. Observe that in the *Pay* process (Algorithm 4), from the recipient $y$'s point of view, the transaction $tx$ sent by $x$ represents an income, and its cost is computed at Line 4, in which $y$ executes the transaction set $\Theta$. For instance, in the third example above where Alice receives tokens from Dave via transaction $tx_{da}$, if $tx_{da}$ went through the *Pay* process, then Alice had already known the cost of $tx_{da}$ (*i.e.*, the gas cost of $tx_{cy}$) since she had executed $tx_{yd}$ during the verification of $tx_{da}$.

In case the user does not already know the cost of an income transaction, she needs to explicitly compute this cost. Further, Line 4 of the *Pay* process also incurs much computation. While these operations could be done by the user herself, *e.g.*, using rental virtual machines from a cloud platform, it is often more continent for the user to outsource such tasks to a professional accountant through the same blockchain, detailed next.

**On-chain accountants.** ANH envisions that third-party professional *accountants* help users perform computational accounting tasks, including determining the income costs and (selectively) executing a given set $\Theta$ of transactions during the *Pay* process described in Algorithm 4. Such a service can achieve economy of scale by serving multiple users, especially when their transactions share common dependencies in the world state, whose results can be cached to improve efficiency. Further, the accountant may possess the expertise and / or specialized algorithms for reducing the cost of performing computational accounting tasks, which are not commonly available to ordinary users.

Such an accountant can be a trusted party, akin to real-world accounting firms and notary public services, who are legally bound to produce correct results. Note that if the accountant provides incorrect results to the client, he/she risks being caught lying, which happens when an honest party executes the same transactions and verifies the accountant's answers with the true results. The accounting service can also be performed via the same blockchain, referred an *on-chain accountant*. There are various ways to implement such an accountant through smart contracts. One way is to require that the accountant deposit a considerable amount of tokens to a smart contract that contains an *Oath-of-Correctness* function, as shown below.

---
*Algorithm 5: Oath-of-Correctness(Q, r, p)*
// Input: Q: a query about a given world state at a given time.
//           r: a value that is claimed to be the result of Q.
//           p: penalty amount for a falsified result r.
1. If $r \neq$ the result of query Q, then:
2.     Transfer p tokens to a blackhole account. // slash

---

Basically, the *Oath-of-Correctness* function in the smart contract takes as input (i) a user's query Q, whose result can be uniquely determined by past transactions on the blockchain, and (ii) the accountant claimed result r of Q. Once the account finishes the computations requested by the client, it calls *Oath-of-Correctness* with the results. The client can then directly read r from the *Oath-of-Correctness* transaction description, without executing any transaction on the blockchain. If the accountant falsifies r, then he/she gets *slashed* (Line 2) and loses a certain amount of tokens (specified by p) from the deposit as a penalty, which goes to a blackhole account that no one knows the private key, *e.g.*, 0x0. Alternatively, the penalty tokens can be paid to the client as compensation.

Such an oath discourages an honest accountant from returning incorrect results. Note that it is computationally intensive to know whether the oath contract has enough deposit for the slash operation to take effect (Line 2). To establish that, the oath smart contract account can be periodically audited by another accountant. The frequency of such auditing depends on the trustworthiness of the accountant (*e.g.*, by the amount of deposit and her history of serving other clients), which might be automated via machine learning.

It is worth mentioning that instead of setting up an oath smart contract at the accountant (*e.g.*, Algorithm 5), it is also possible to set up a *bounty* smart contract at the client, which pays the accountant only if the answer is correct. When the accountant is a professional service and the client is an ordinary user, it might appear more natural to set up the smart contract at the accountant's side. The idea of a client-side bounty smart contract is revisited later in Section III.

### D. Collecting Transaction Fees

So far, our discussion has not addressed an important issue: the collection of transaction fees by validators. Unlike traditional blockchains (*e.g.*, Bitcoin [14]) in which the transaction fee can (theoretically) be zero and the miners are paid by the platform, in ANH, each transaction must incur a fee, in order to prevent the *Tx-DoS* attack described in Algorithm 1. Further, to effectively defend against *Tx-DoS*, the transaction fee payment for any transaction *tx* must be immediately and explicitly confirmed by the block validators during the consensus protocol; in particular, if the sender's account does not have sufficient balance to pay the fee for *tx*, then *tx* should be ignored and not added to a new block. However, paying transaction fees with the *Pay* process (Algorithm 4) is clearly infeasible, since the consensus protocol needs to be performed with stringent time constraints.

To address this problem, ANH imposes the following rules:
1. Transaction fees must be paid with zero-cost income (Definition 3).
2. The transaction sender must pay the maximum possible gas costs specified by the *gas limit* of the transaction. If the transaction completes without reaching the gas limit, the remaining paid gas is returned to the sender as an income of the sender's account.

When an account does not sufficient zero-cost income to cover the transaction fee, its transactions are immediately rejected by the block validators during the consensus protocol. The following lemma clarified the source of zero-cost incomes, whose proof follows from Definition 3, and is omitted for brevity.

*Lemma 2 (Source of Zero-Cost Income).* Each zero-cost income w.r.t to an account *x* can be traced back to a source of tokens directly provided by the blockchain platform, *e.g.*, the genesis block, through a sequence of direct user-to-user token transfers ending at *x*.

When a user runs low on zero-cost income, it may use the service of a professional *money changer* service, discussed in Section II-E.

**Fee structure.** Next we clarify the necessary components of the transaction fee. As explained above, there must be a minimum fee for each transaction, to deter the *Tx-DoS* attack (Algorithm 1). Further, to prevent infinite loops, there needs to be a minimum gas charge, which can be set far lower than traditional blockchain platforms, since transactions on ANH are executed on a need-to-run basis, and might never be executed, *e.g.*, *Exec-DoS* and *Targeted-Exec-DoS* presented in Section II-B.

In existing platforms, the gas costs are designed without consideration for DNN training, which is typically performed on GPUs or TPUs. For instance, the multiplication of two large matrices can be done efficiently using hardware-accelerated

API calls (*e.g.*, Nvidia CUDA[8]); therefore, the gas cost of a matrix multiplication operation should be far lower than the sum of the gas cost for the underlying elementary operations such as multiplications and additions. Note that the gas cost of different operations also affects DNN architecture design. For instance, on GPUs, a 3x3 convolution costs significantly less than 9 1x1 convolutions, which has been exploited in ConvNet designs such as ShuffleNet [13]. Hence, the gas cost should agree with the hardware benchmark results to provide the correct incentives for efficient DNN architecture design.

Finally, existing platforms also incur rather large gas costs for memory consumption, since each node is required to maintain the entire world state. In ANH, memory gas cost can be significantly reduced for similar reasons as computational gas cost. Meanwhile, large objects, such as weights of a trained DNN, should be placed in secondary decentralized storage, *e.g.*, via IPFS [2], rather than in memory.

### E. Discussions

**Token fungibility and economics.** Roughly speaking, fungibility means every token should hold the same value, regardless of its history. ANH trades token fungibility with computational efficiency on the blockchain platform, by deferring smart contract computations to payment time (Algorithm 4), which leads to the concept of computation cost of income (Definition 3). This concept may appear esoteric at first glance; a closer look reveals that income cost in fact applies to the physical (*i.e.*, off-chain) world too. In particular, any sizable amount of money, regardless of the underlying currency type or technology, is hardly fungible, due to various legal constraints. For instance, different income items can be taxed at different rates, depending on the type (*e.g.*, salary, capital gain, inheritance, divorce payments), source (*e.g.*, domestic, foreign), *etc*. In this sense, a common task that accountants in the physical world perform is to compute solutions to the optimization problem that minimizes the cost (*e.g.*, tax obligations) of money.

In ANH, the accounting cost of money increases monotonously over time, which may cause problems both on the microeconomic and macroeconomic levels. On the microeconomic level, an income item whose accounting cost exceeds its value is essentially worthless to the user. Note that the accounting cost varies depending on the user: to an ordinary individual, the cost can be rather high since the user needs to purchase retail services from an accountant; to a large financial institution with its own data centers and accounting team, the cost can be significantly lower. Hence, one way to mitigate the problem is through dedicated *money changer* services, which exchange high-cost to low-cost (or even zero-cost) money at a fee. The design of such a service is outside the scope of this paper, and is left as future work.

On the macroeconomic level, the total accounting cost of all tokens (which is initially zero in the genesis block) increases with every smart contract function call, and can never decrease. Meanwhile, the blockchain platform requires low-cost money to function, *e.g.*, transaction fee payment must be done with zero-cost income, as explained in Section II-D. This problem needs to be addressed on the blockchain platform level through governance rules. For instance, a blockchain may introduce sources of zero-cost money after the genesis block (*e.g.*, payments to block validators, airdrops, *etc.*), which resembles a central bank injecting liquidity to the economy.

Another way is to selectively adopt the traditional blockchain approach in ANH, leading to a hybrid system. In particular, a rich account may present a high-cost income item to all validators in the blockchain, who confirm the income by executing the relevant transactions locally, reaching consensus over all relevant world states, and agreeing to replace the expensive income item (*e.g.*, by dumping it to a blackhole account) with newly minted zero-cost tokens issued by the platform. Such measures are left to the implementation of the specific blockchain platform adopting the ANH designs.

**ANH vs. sharding.** ANH reduces total smart contract computation costs through lazy, on-demand execution. In the cryptocurrency community, the hope for fast transaction execution has been mostly placed on *sharding* (*e.g.*, [12]) which processes transactions in parallel. Note that there is a limit to parallelization: for instance, even with sharding, a slow operation (such as DNN training) that takes hours still cannot be done within a single transaction, if the platform expects to produce a block, say, every 10 minutes.

Further, the level of parallelism is also limited by the dependencies between smart contract function calls and world state variables. For example, in SeaLevel [23], each smart contract transaction is required to declare all its dependent variables, so that the execution engine can maximize parallelism accordingly. For this reason, some of the challenges faced by ANH also apply similarly to sharding. For instance, in a targeted DoS attack of ANH (Algorithm 3), the adversary intentionally introduces dependencies to a world state variable, *i.e.*, balance of the victim account $x$. In sharded executions, the adversary (*e.g.*, a competing DeFi service) may similarly aim to slow down a smart contract function (*e.g.*, a victim DeFi service), by issuing transactions that depend upon the victim's variables and a large number of other variables, to reduce the victim's transaction parallelism.

Finally, observe that ANH can be viewed as an extreme form of sharding, in the sense different transactions are usually executed by different parties, as transactions are run on a need-to-run basis. Unlike sharding, however, the number of runs for each transaction can vary in ANH, which can be as low as zero.

**Impact on smart contract design.** ANH requires special considerations in smart contract design. For instance, the concept of minimal accounting query (MAQ) needs to be defined based on the application's requirements, as explained in Section II-B. Further, unlike user accounts, both expenses and income of a smart contract can be complex and computationally intensive. For instance, a smart contract may contain a function that sends out tokens when it receives an income, or upon reaching certain conditions represented by the contract's storage variables. Such an expense can have complex dependencies. To make the smart contract usable, the application designers and developers should aim to reduce transaction dependencies and minimize execution costs for important functionalities. Such designs are application-dependent, and are outside the scope of this paper.

---

[8] *https://developer.nvidia.com/cuda-zone*

Lastly, for smart contract applications that have time constraints, *e.g.*, a decentralized exchange (DEX), ANH recommends limiting the total computational cost of specific types of transactions (including the cost of all dependent transactions), similar to the case that transaction fees must be paid with zero-cost incomes. Details are left as future work.

### III. Non-Deterministic Smart Contracts

Allowing a smart contract function to return non-deterministic results in ANH opens new possibilities for attacks. For instance, consider a trading bot, which buys or sells tokens guided by a DNN. Recall that in ANH, transactions are executed after they are committed to the blockchain, at which time the user has gained new information that was not available at the time the transaction was submitted. Based on this new information (*e.g.*, token price movements), the user might attempt to argue that the non-deterministic DNN returned favorable results (bought tokens shoes price turned out to rise). To avoid this, ANH needs effective and efficient mechanisms to establish verifiable transaction results in the presence of non-deterministic functions.

*The rest of this section is intentionally left empty. The contents will appear in the next version of this preprint.*